\documentstyle[preprint,aps]{revtex}
\begin{document}

\draft
\title
{\bf Light-emitting current of electrically driven single-photon
sources}

\author{David M.-T. Kuo}
\address{Department of Electrical Engineering, National Central
University, Chung-Li, Taiwan, 320, Republic of China}
%\address
%{$^2$Department of Physics\\
%University of Illinois at Urbana-Champaign, Urbana, Illinois
%61801}

%\end{center}
\date{\today}
\maketitle

\begin{abstract}
Time-dependent tunnelling current arising from the electron-hole
recombination of the exciton state is theoretically studied using
the nonequilibrium Green's function technique and the Anderson
model with two energy levels. The charge conservation and gauge
invariance are satisfied in the tunnelling current. Apart from the
classical capacitive charging and discharging behavior,
interesting oscillations superimpose on the tunnelling current for
the applied rectangular pulse voltages.
\end{abstract}

\newpage
\section{Introduction}
Recently, much effort has been devoted to the study of
single-photon sources (SPSs) made from quantum dots (QDs), which
have potential applications in quantum cryptography and quantum
computing$^{1-3)}$. The second order correlation function of light
emitted from such devices must be measured to determine their
functionality as SPSs, that is, the light source antibunching
feature must be examined. The antibunching features of SPSs were
demonstrated in Ref. [1,2], where electrons and holes in the QD
are excited by optical pumping. Nevertheless, only a few
experiments have employed electrical pumping is to demonstrate the
antibunching behavior of SPSs.From a practical point of view, it
is more efficient to use the electrically driven SPSs. A prototype
SPS made from individual QDs embedded in a semiconductor p-n
junction was suggested by Imamoglu and Yamamoto$^{4}$. In addition
to antibunching, enhancing the spontaneous emission rate is
necessary, because it is crucial to SPS performance.

For small size QDs, the strong three dimensional confinement
effect creates large energy separations among the low-lying
confined levels. Consequently, it is possible to inject electrons
and holes into the corresponding ground state energy levels of a
single QD and generate photons via electron-hole recombination in
the exciton, positive/negative trion or biexciton states in the
QD. To fabricate a single of nanometer-sized QD  at a specific
location is one of the most challenging processes in the
realization of electrically driven SPSs. Self-assembled quantum
dots (SAQDs) combined with a selective formation method may
overcome such a difficulty$^{5}$. Even though, some theoretical
studies have been devoted to electrically driven SPS
devices,$^{6-8}$ the dynamic properties of an electrically driven
SPS are still not clear. In ref. 4, the Monte Carlo method was
used to simulate junction dynamics, where time-dependent junction
voltage is calculated. Other studies emphasized steady state
characteristics such as the exciton assistant tunnelling
current$^{6}$, Purcell effect on the tunnelling current$^{7}$ and
electrode effects on the exciton complexes$^{8}$.

The main purpose of this study is to investigate the
time-dependent tunnelling current (or spontaneous emission rate)
arising from electron-hole recombination of exciton state by using
nonequilibrium Green's function technique$^{9-11}$, which has been
applied to several different systems.$^{12-15}$. This study
attempts to clarify how the carrier tunnelling process, applied
modulation voltage and temperature influence the spontaneous
emission rate in an electrically driven SPS. The schematic band
diagram describing the studied system is shown in Fig. 1. Only the
ground states of conduction and valence band of the QD are
considered. The energy level of $E_2$ is 20~meV below the Fermi
energy level of left electrode $E_{F,e}=50$~meV.  On the other
hand, the exciton state denoted by $E_{ex}$ is 15~meV above the
Fermi energy of right electrode $E_{F,h}=50$~meV. $\Gamma_{2(1)}$
denotes the tunnelling rate from the QD to the left (right)
electrode. The emitted photons will be observed when the periodic
square bumplike modulation v(t) added into the right electrode
injects holes into the exciton resonant level.

\section{Hamiltonian}
An Anderson model with two energy levels in terms of electron
picture is used to describe the system as shown in Fig. 1,

\begin{eqnarray}
H&=&\sum_{k} \epsilon_{k,L} a^{\dagger}_{k}a_{k}+\sum_{k}
\epsilon_{k,R} b^{\dagger}_{k}b_{k}+ \sum_{i=1,2} \epsilon_i
d^{\dagger}_{i} d_{i}\\ \nonumber &+ &\lambda_{12} d^{\dagger}_{1}
d_{2}+ \lambda^{*}_{21} d^{\dagger}_{2} d_{1} +\sum_{k,1}
t_{k,1}b^{\dagger}_{k}d_{1}+h.c\\ \nonumber
&+&\sum_{k,2}t^{\dagger}_{k,2}a^{\dagger}_{k}d_{2}+h.c
\end{eqnarray}
where $a^{\dagger}_{k} (a_{k})$ and $b^{\dagger}_{k} (b_{k})$
create (destroy) an electron of momentum $k$ in the left and right
system electrodes, respectively. The free electron model is used
in the electrodes in which electrons have frequency-dependent
energies $\epsilon_{k,L}=\varepsilon_k-\omega/2$ and
$\epsilon_{k,R}=\varepsilon_k+\omega/2+v(t)$. Time-dependent
modulation v(t) denotes the time-dependent applied voltage in the
right electrode. $d^{\dagger}_{i}$ ($d_{i}$) creates (destroys) an
electron inside the QD with orbital energy
$\epsilon_i=E_i-(-1)^i\omega/2$. In this study $i=1$ and $i=2$
represent, respectively, the ground states of valence band and
conduction band of individual QDs. The fourth and fifth terms
describe the coupling of the QD with electromagnetic field of
frequency $\omega$. $\lambda=-\mu_r {\cal E}$ is the Rabi
frequency, where $\mu_r=<f|r|i>$ is the matrix element for the
optical transition and ${\cal E}$ is the electric field per
photon. $t_{k,i}$ describes the coupling between the band states
of electrodes and energy levels of the QD. Note that a unitary
transformation, $S(t)=exp^{i\omega t/2(\sum_k
(b^{\dagger}_{k}b_{k}-a^{\dagger}_{k}a_{k})+d^{\dagger}_{1}
d_{1}-d^{\dagger}_{2} d_{2})}$, has been used to obtain Eq. (1)
via
\[
H=S^{-1}H(t)S-iS^{-1}\frac{\partial}{\partial t}S.
\]
For small semiconductor QDs, the particle correlation is strong.
This indicates that the intralevel Coulomb interactions of the QD,
$U_{11}$ and $U_{22}$, can not be ignored. In order to simplify
problem, it is assumed that a regime of the applied voltage can
not overcome the charging energies $U_{11}$ and $U_{22}$.$^{8}$ To
investigate the exciton assistant process$^{6,8}$, the interlevel
Coulomb interaction $U_{12}$ is taken into account in Eq. (1)
\begin{equation}
H_U=U_{12} d^{\dagger}_1 d_1 d^{\dagger}_2 d_2,
\end{equation}
which is invariant under unitary transformation.

\section{Formalism}
Although a spontaneous emission process of photons is the quantum
effect of an electromagnetic field$^{8}$, the electromagnetic
field is still treated as a semiclassical field in Eq. (1). The
approach detailed in ref.[11] is employed to study the spontaneous
emission process of photons. The optical susceptibility of QDs not
only provides the absorption coefficient and the refractive index,
but also determines the spontaneous emission rate.$^{11}$
Therefore, the optical susceptibility of individual QDs is
calculated using the lesser Green's function defined as
$G^{<}_{i,j}(t,t')=i<d^{\dagger}_{i}(t')d_j(t)>$. Based on
Keldysh's equation, we have

\begin{equation}
G^{<}_{i,j}(t,t)=\int dt_1 \int dt_2
G^{r}_{i,n}(t,t_1)\Sigma^{<}_{n,m}(t_1,t_2)G^{a}_{m,j}(t_2,t),
\end{equation}
where $\Sigma^{<}_{n,m}$, $G^r_{i,n}$ and $G^{a}_{m,j}$ are the
lesser self-energy, the retarded Green's function and the advanced
Green's function, respectively. The Einstein summation index is
used in Eq. (3). The spontaneous emission rate will be suppressed
if tunnelling rates $\Gamma_2$ and $\Gamma_1$ are smaller than
spontaneous emission rate $R_{eh}$. The detailed expression of
$R_{eh}$ is given later. To avoid the suppression of the
spontaneous emission rate, the device shown in Fig. 1 favors the
condition of $\Gamma_2
> \Gamma_1 >> R_{eh}$ or $\Gamma_1 > \Gamma_2 >> R_{eh}$ to
function as a light emitting source.  The condition of $\Gamma_2
> \Gamma_1 >> R_{eh}$, $\lambda_{ij}$ will be regarded as a small
parameter in the comparison with $t_{k,i}$. Consequently, the
lesser self-energy of Eq. (3) is mainly attributed to tunnelling

\begin{equation}
\Sigma^{<}_{i,i}(t,t')=i\int \frac{d\epsilon}{2\pi} \Gamma_i
f_i(\epsilon) e^{-i\epsilon(t-t')},
\end{equation}
where the tunnelling rates are $\Gamma_i=\sum_k t_{k,i}
t^{*}_{k,i} \delta(\epsilon-\varepsilon_k)$, and
$f_i(\epsilon)=1/(e^{\epsilon-\mu_{F,i}}-1)$ is the Fermi
distribution function of the electrodes, in which the chemical
potential is given by $\mu_{F,i}=E_{F,i}-(1)^i\omega/2$. It is
worth noting that energy-independent tunnelling rates are assumed
in Eq. (4) for the sake of simplicity. Inserting Eq. (4) into Eq.
(3),we obtain two off-diagonal lesser Green's functions
\begin{eqnarray}
G^{<}_{2,1}(t,t)&=&i\int \frac{d\epsilon}{2\pi}
\{\Gamma_{1}f_1(\epsilon) A^r_{21}(\epsilon,t)
A^a_{11}(\epsilon,t)\\ \nonumber &+&\Gamma_2 f_2(\epsilon)
A^{r}_{22}(\epsilon,t) A^{a}_{21}(\epsilon,t)\},
\end{eqnarray}
and
\begin{eqnarray}
G^{<}_{1,2}(t,t)&=&i\int \frac{d\epsilon}{2\pi}
\{\Gamma_{2}f_2(\epsilon) A^r_{12}(\epsilon,t)
A^a_{22}(\epsilon,t)\\ \nonumber &+&\Gamma_1 f_1(\epsilon)
A^{r}_{11}(\epsilon,t) A^{a}_{12}(\epsilon,t)\},
\end{eqnarray}
as well as two diagonal lesser Green's functions

\begin{eqnarray}
G^{<}_{1,1}(t,t)&=&i\int \frac{d\epsilon}{2\pi}
\{\Gamma_{1}f_1(\epsilon) A^r_{11}(\epsilon,t)
A^a_{11}(\epsilon,t)\\ \nonumber &+&\Gamma_2 f_2(\epsilon)
A^{r}_{12}(\epsilon,t) A^{a}_{21}(\epsilon,t)\},
\end{eqnarray}
and
\begin{eqnarray}
G^{<}_{2,2}(t,t)&=&i\int \frac{d\epsilon}{2\pi}
\{\Gamma_{2}f_2(\epsilon) A^r_{22}(\epsilon,t)
A^a_{22}(\epsilon,t)\\ \nonumber &+&\Gamma_1 f_1(\epsilon)
A^{r}_{21}(\epsilon,t) A^{a}_{12}(\epsilon,t)\},
\end{eqnarray}
where $\int dt_1 e^{i\epsilon(t-t_1)}G^r_{ij}(t,t_1)
=A^r_{ij}(\epsilon,t)$ and $\int dt_1
e^{-i\epsilon(t-t_1)}G^a_{ij}(t,t_1) =A^a_{ij}(\epsilon,t)$. The
expression of $A^{r(a)}_{ij}(\epsilon,t)$ depends on the detailed
form of retarded and advanced Green's functions. Eqs. (5) and (6)
determine the optical susceptibility of QDs. Eqs. (7) and (8)
determine the electron occupation numbers. Once $G^{<}_{21}(t,t)$
is determined, $G^{<}_{12}(t,t)$ is also obtained as a result of
$G^{<}_{21}(t,t)=(G^{<}_{12}(t,t))^{\dagger}$. To solve
$G^{<}_{21}(t,t)$, some approximations are considered in the
following derivation due to the complicated Hamiltonian of Eq.
(1). Because the energy level of i=2 emerges into the band width
of the left electrode and $\Gamma_2$ is larger than $\Gamma_1$,
the retarded and advanced Green's functions for i=2 can be
regarded as a steady state solution. That is $G^{r(a)}_{22}(t,t')=
\mp~i\theta(t-t')e^{-i(\epsilon_1\mp\Gamma_2/2)(t-t')}$, which
gives
$A^{r(a)}_{22}(\epsilon,t)=1/(\epsilon-\epsilon_2\pm~i\Gamma_2/2)$.
As for $G^{r(a)}_{i,j}(t,t')$, the off-diagonal Green's function
($i\neq j$) are solved using Dyson's equation,

\begin{equation}
G^{r(a)}_{ij}(t,t')=\int dt_1 G^{r(a)}_{i,i}(t,t_1)\lambda_{i,j}
G^{r(a)}_{j,j}(t_1,t').
\end{equation}
Due to $\lambda_{i,j} < t_{k,i}$ and $G^{r(a)}_{ij}(t,t')$ in
terms of the first order parameter $\lambda$, the Green's
functions $G^{r(a)}_{ii}$ in Eq. (9) are $\lambda$-independent
functions. Consequently, $A^{r(a)}_{21}(\epsilon,t)=\lambda_{21}
A^{r(a)}_{22}(\epsilon) A^{r(a)}_{11}(\epsilon,t)$ and  Eq. (5) is
rewritten as

\begin{eqnarray}
G^{<}_{2,1}(t,t)&=& \lambda_{21} \int \frac{d\epsilon}{2\pi}
\{f_1(\epsilon) \Gamma_1 |A^r_{11}(\epsilon,t)|^2 A^{r}_{22}(\epsilon)\\
\nonumber &+& f_2(\epsilon) \Gamma_2 |A^{r}_{22}(\epsilon)|^2
A^{a}_{11}(\epsilon,t)\}.
\end{eqnarray}
To obtain the spontaneous rate resulting from the imaginary part
of $G^{<}_{2,1}(t,t)$, the imaginary part on the both sides of Eq.
(10) is used
\begin{eqnarray}
& &ImG^{<}_{2,1}(t,t) \nonumber \\&=& \lambda_{21} \int
\frac{d\epsilon}{2\pi}
\{f_1(\epsilon) \Gamma_1 |A^r_{11}(\epsilon,t)|^2 \frac{1}{2}[A^{r}_{22}(\epsilon)-A^{a}_{22}(\epsilon)]\\
\nonumber &+& f_2(\epsilon) \Gamma_2 |A^{r}_{22}(\epsilon)|^2
\frac{1}{2}[A^{a}_{11}(\epsilon,t)-A^{r}_{11}(\epsilon,t)]\}.
\end{eqnarray}
Employing $G^{r}_i-G^{a}_i=G^{>}_i-G^{<}_i$, where the greater
Green's function is given by
\begin{equation}
G^{>}_{i,j}(t,t)=\int dt_1 \int dt_2 G^{r}_{i,n}(t,t_1)
\Sigma^{>}_{n,m}(t_1,t_2)G^{a}_{m,j}(t_2,t)
\end{equation}
with the greater self-energy
\begin{equation}
\Sigma^{>}_{i,i}(t,t')=i\int \frac{d\epsilon}{2\pi} \Gamma_i
(1-f_i(\epsilon)) e^{-i\epsilon(t-t')},
\end{equation}
Eq. (11) can be simplified as
\begin{eqnarray}
& &ImG^{<}_{e,h}(t,t)\\ \nonumber &=& -\lambda_{eh} \int
\frac{d\epsilon}{2\pi} \{(1-f_e(\epsilon)-f_h(\epsilon)) \Gamma_h
|A^r_{hh}(\epsilon,t)|^2 \Gamma_e|A^{a}_{ee}(\epsilon)|^2\}.
\end{eqnarray}

In Eq. (14) the electron and hole picture are employed to label
i=2 and i=1, respectively.  Note that Eq. (14) contains a factor
of $(1-f_e(\epsilon)-f_h(\epsilon))$. It is always possible to
write $\lambda_{eh} ImG^{<}_{e,h}(t,t)={\cal X}_e(\omega,t)-{\cal
X}_a(\omega,t)$, where ${\cal X}_a(\omega,t)$ and ${\cal
X}_e(\omega,t)$ are, respectively, in proportion to
$(1-f_e(\epsilon))(1-f_h(\epsilon))$ and
$f_e(\epsilon)f_h(\epsilon)$. ${\cal X}_a(\omega,t)$ and ${\cal
X}_e(\omega,t)$ denote the absorption and emission spectra,
respectively. We only focus on the emission spectrum, which was
frequently reported in experiments,
\[ {\cal
X}_e(\omega,t)=\lambda^2_{eh} \int \frac{d\epsilon}{2\pi}
\{f_e(\epsilon)f_h(\epsilon) \Gamma_h |A^r_{hh}(\epsilon,t)|^2
\Gamma_e|A^{a}_{ee}(\epsilon)|^2\}.
\]
By reference to ref. 11, the current arising from the
electron-hole recombination of the exciton state may be written as

\begin{eqnarray}
& &J_{sp}(t)\\ \nonumber &=&e \alpha \int d\omega~\omega^3 \int
\frac{d\epsilon}{\pi^2} f_e(\epsilon)f_h(\epsilon) \Gamma_h
|A^r_{hh}(\epsilon,t)|^2 Im[A^a_{ee}(\epsilon)]
\end{eqnarray}
with  $\alpha=4n^3_r \mu^2_r/(6c^3\hbar^3\epsilon_0)$, where $n_r$
and $\epsilon_0$ are the refractive index and static dielectric
constant of system, respectively. e denotes the electron charge.
Note that Eq. (15) uses
$\Gamma_e|A^{a}_{ee}(\epsilon)|^2=2~Im[A^a_{ee}(\epsilon)]$ which
is valid only for the steady state. It should be noted that
$A^{r}_{hh}(\epsilon,t)$ satisfies the condition of
$\Gamma_h|A^{r}_{hh}(\epsilon,t)|^2=
-2~Im[A^r_{hh}(\epsilon,t)]-\frac{d|A^{r}_{hh}(\epsilon,t)|^2}{dt}$
resulting from total charge conservation and gauge
invariance$^{16}$. $\rho(\omega)=\omega^2$ of Eq. (15) arises from
the density of states of photons. $J_{sp}(t)$ is determined by the
time-dependent interband joint density of states and the factors
of $f_e(\epsilon) f_h(\epsilon)$. According to the expressions of
${\cal X}_e(\omega,t)$ and $J_{sp}$, we prove that the intensity
of emission spectrum is linear variation with respect to
$J_{sp}$.$^{3}$

To solve the spectral function of holes $A^{r}_{hh}(\epsilon,t)$,
the retarded Green's function of holes $G^{r}_{hh}(t,t_1)$ is
derived to obtain
\begin{equation}
G^{r}_{hh}(t,t_1)=(1-N_e)g^{r}_h(\epsilon_h,t,t_1)+N_e
g^{r}_h(\epsilon_h+U_{eh},t,t_1)
\end{equation}
with
\begin{equation}
g^r_h(\epsilon_h,t,t_1)=-i\theta(t-t_1)e^{-i(\epsilon_h-i\Gamma_h/2)(t-t_1)-i\int^t_{t_1}dt_2v(t_2)},
\end{equation}
where $\epsilon_h=-E_h+\frac{\omega}{2}$, and $N_e$ is the
electron occupation number at steady state. Two branches exist in
Eq. (16) ; one corresponds to the resonant energy level of $E_h$
with a weight of $(1-N_e)$, and the other corresponds to the
exciton resonant level of $E_{ex}=E_h-U_{eh}$ with a weight of
$N_e$. Consequently, holes injected into the energy levels of QDs
depend not only on the applied voltage of the right electrode, but
also on the electron occupation number $N_e$, which is given by
$N_e=\int \frac{d\varepsilon}{\pi} f_e(\varepsilon)
Im\frac{1}{\varepsilon-E_e-i\Gamma_e/2}$. As long as the applied
voltage is insufficient to inject holes into the resonant energy
level of $E_h$, it is only necessary to consider the exciton
branch of $N_e g^{r}_h(\epsilon_h+U_{eh},t,t_1)$ in Eq. (16). The
detailed expression of $G^{r}_{hh}(t,t_1)$ depends on the applied
voltage v(t), considering a rectangular pulse with the duration
time of $\Delta s$ and amplitude $\Delta$ as v(t). For $t_0 \le t
< t_0+\Delta s$, according to Eq. (17),
\begin{eqnarray}
&& A^r_h(\epsilon,t)\\ \nonumber &=&
\frac{N_e}{\epsilon-\epsilon_h+i\Gamma_h/2} (
1+\frac{\Delta}{\epsilon-\epsilon_h-\Delta+i\Gamma_h/2}
\\ \nonumber& &(1-e^{i(\epsilon-\epsilon_h-\Delta+i\Gamma_h/2)(t-t_0)})
).
\end{eqnarray}
For $t \ge \Delta s$,
\begin{eqnarray}
&& A^r_h(\epsilon,t)\\ \nonumber &=&
\frac{N_e}{\epsilon-\epsilon_h-\Delta+i\Gamma_h/2} (
1+\frac{\Delta}{\epsilon-\epsilon_h+i\Gamma_h/2}
\\ \nonumber& &(1-e^{i(\epsilon-\epsilon_h+i\Gamma_h/2)(t-\Delta s)}
(1-e^{i(\epsilon-\epsilon_h-\Delta+i\Gamma_h/2)\Delta s}) ).
\end{eqnarray}
The time-dependent tunnelling current associated with the
spontaneous radiative transition in individual quantum dots is
next discussed based on Eqs. (15), (18) and (19).

\section{Results and discussion}
Due to current conservation, the time-dependent tunnelling current
is given by $J(t)=J_{sp}(t)=e\Gamma_s(t)$, where $\Gamma_s(t)$
denotes the time-dependent spontaneous emission rate. Owing to
small tunnelling rates ($\Gamma_e/E_{F,e}<< 1$ and $\Gamma_h
/E_{F,h} << 1$), Eq. (15) can be reduced to
\begin{equation}
J_{sp}(t)=2e*R_{eh} f_e(T) A_{h}(T,t),
\end{equation}
with
\begin{equation}
A_h(t)=\int \frac{d\varepsilon}{\pi} f_h(\varepsilon)
\Gamma_h|A^r_h(\varepsilon,t)|^2.
\end{equation}
In Eq. (20) we define the time-independent spontaneous emission
rate $R_{eh}=\alpha~\Omega^3_{ex}$, where
$\Omega_{ex}=E_g+E_e+E_h-U_{eh}$. According to Eq. (20), the
time-dependent feature of $J_{sp}(t)$ is determined by the hole
occupation number of Eq. (21).

To reveal the time-dependent behavior of $J_{sp}(t)$, Eq. (20) is
solved numerically, and shows $J_{sp}(t)$ for various duration
times at zero temperature in Fig. 2 for the tunnelling rate
$\Gamma_h= 0.5~$meV and $\Delta=20$ meV; the solid line ($\Delta
s=t_0$), the dashed line ($\Delta s= 3~t_0$) and the dotted line
($\Delta s= 5~t_0$). The current almost reaches saturation at
$t=5~t_0$. Apart from the classical capacitive charging and
discharging behavior (exponential growth and decay)$^{17}$, the
interesting oscillations superimpose on the tunnelling current not
only for $t \leq \Delta s$ but also for $t
> \Delta s$. However, the amplitude of oscillation for $t \leq
\Delta s$ is small. In particular, the oscillation period for $t
>\Delta s$ does not depend on the magnitude of duration time.
The oscillatory current is yielded from the particle coherent
tunnelling between the electrodes and the QD. Such a coherent
tunnelling was also pointed out as the mechanism of quantum
ringing for electrons tunnelling current through the single dot
embedded in an n-i-n structure$^{18}$. In ref. 18, oscillatory
current is not observed as $t > \Delta s$. According to the result
of quantum interference between the outgoing wave (leaving the
QDs) and the wave reflected from the barrier, this oscillatory
behavior could be observed in the discharging process$^{19}$.

To examine if the oscillation currents shown in Fig. 2 depend on
applied bias strength, Figure 3 illustrates $J_{sp}(t)$ for
different applied biases with duration time $\Delta s=3~t_0$ at
zero temperature and $\Gamma_h=0.5~meV$. The solid line denotes
$\Delta=15~meV$, and the dashed and dotted lines denote,
$\Delta=20~meV$ and $\Delta=25~meV$, respectively. For amplitude
$\Delta=15~meV$, the Fermi energy of the right electrode just
reaches the alignment with the resonant exciton level. $J_{sp}(t)$
displays a strong oscillation characteristic. If the pulse
amplitude is increased to $\Delta=20 meV$, the resonant exciton
level is covered by the right electrode reservoir, and therefore
the magnitude of $J_{sp}(t)$ increases. When the pulse amplitude
to $\Delta=25~meV$, $J_{sp}(t)$ becomes saturated . The results
shown in Fig. 3 clearly indicate that the oscillation feature of
$J_{sp}(t)$ depends on the magnitude of $\Delta$. Due to the
emitted photon numbers in proportion to $J_{sp}(t)$, the features
of emitted photon numbers as functions of time will correspond to
the results shown in Fig. 3.

In Figs. (2) and (3) hole tunnelling rate is set as
$\Gamma_h=0.5~meV$, it is interesting to understand how the
tunnelling rate influences $J_{sp}(t)$. $J_{sp}(t)$ for different
hole tunnelling rates is shown in Fig. 4; the solid line
($\Gamma_h=0.5$~meV), the dashed line ($\Gamma_h=0.75$~meV) and
the dotted line($\Gamma_h=1$~meV). Increasing the tunnelling rate,
$J_{sp}(t)$ reaches exponential growth quickly. This also
indicates that electrically driven SPSs can quickly reach the
maximum photon emission efficiency within a shorter operating time
compared to optically driven SPSs with a phonon bottleneck. To
clarify, the relation between the photon number $N_s$ and
$J_{sp}(t)$ should be constructed. Because $\frac{dN_{s}(t)}{dt}=
\Gamma_{sp}(t)$, the time-dependent photon number
$N_{s}(t)=\int^{t}_{0}dt~\Gamma_{sp}(t)$ is obtained. For the
step-like pulse ($\Delta s= \infty$,), $A^r_h(\epsilon,t)$ is
given by Eq. (18). Due to the small oscillatory amplitude in the
charging process, Eq. (18) is approximated as $A_h(t)= f_h(T)
(1-exp^{-\Gamma_h t})$ for $t \ge 0$. Consequently, $N_s(t)=R_{eh}
f_e(T)f_h(T) t (1+(exp^{-\Gamma_h t}-1)/(\Gamma_h t))$. We found
that $N_s(t)/t= R_{eh} f_e(T) f_h(T)$ as $\Gamma_h t >> 1$. This
indicates that a higher tunnelling rate allows an electrically
driven SPS to quickly reach the maximum photon emission
efficiency.

Finally, the finite temperature effect on $J_{sp}(t)$ is shown in
Fig. 5; the solid line ($k_BT=0$), the dashed line ($k_BT=1$~meV),
the dotted line ($k_BT=2$~meV) and the dot-dashed line
($k_BT=3$~meV). Fig. 5 shows that the spontaneous photon emission
rate $\Gamma_{sp}(t)$ is suppressed by temperature effects
resulting from a factor of $f_e(T) f_h(T)$ and $N^2_e(T)$ (see
Eqs. (18) and (19)). $N^2_e(T)$ is yielded by the electron-hole
interaction. From Fig. 2 to Fig. 5, $J_{sp}(t)$ is in units of
$J_0=2e~R_{eh}$. For InAs QDs, the typical value of $R_{eh}$ is on
the order of $1~\mu eV (1/ns)$.$^{20}$ This indicates that
$J_{sp}$ results from spontaneous photon emission on the order of
nA, which can be readily measured.

\section{Summary}
The expression of tunnelling current $J_{sp}(t)$ arising from
spontaneous radiative transition in individual quantum dots is
obtained by using the nonequilibrium Green's function method.
$J_{sp}(t)$ is found to be functions of the spontaneous emission
rate $R_{eh}$, the tunnelling rates $\Gamma_h$($\Gamma_e$) and the
factor of $f_e(T) f_h(T)$. In addition to the exponential growth
and decay features corresponding to the charging and discharging
process, the oscillatory behavior of $J_{sp}(t)$ as a function of
time is observed as a result of particle coherent tunnelling
between the electrodes and QDs, which is suppressed by the
temperature effects.

In this study, resonant tunnelling carriers are used to yield the
triggered single-photon sources, in contrast to the captured
carriers typically used in an optically driven SPS.$^{21,22}$ Due
to the phonon bottleneck effect$^{23}$, it is predicted that the
QD capture rate of electrons will be low. This could reduce the
performance of SPS devices, which use captured carriers as the
source of single-photon generations. Using carriers injected via
the resonant tunnelling process can prevent such problems.

{\bf ACKNOWLEDGMENTS}

This work was supported by National Science Council of Republic of
China Contract Nos. NSC 94-2215-E-008-027 and NSC
94-2120-M-008-002.

%\section{Appendix A}

\mbox{}

\newpage

{\bf Figure Captions}

Fig. 1. The schematic band diagram for the single quantum dot
embedded in a p-i-n junction. The ground state energy level of
electrons is $20~meV$ below the Fermi-energy of left-electrode,
and the exciton state for holes is 15~meV above the Fermi-energy
of right-electrode. The periodic square bumplike modulation v(t)
supplies holes into the quantum dot.

Fig. 2. Time-dependent current arising from the electron-hole
recombination $J_{sp}(t)$ for different duration times of applied
rectangular pulse voltage with amplitude $\Delta=20$~meV at zero
temperature. Current is given in units of $J_0=2e~R_{eh}$, where
$R_{eh}$ denotes the time and temperature-independent spontaneous
emission rate, and time is in units of $t_0=\hbar/meV$.

Fig. 3. Time-dependent current arising from the electron-hole
recombination $J_{sp}(t)$ for different applied voltages with
duration time $\Delta s=3~t_0$ at zero temperature. Current is
given in units of $J_0=2e~R_{eh}$, where $R_{eh}$ denotes the time
and temperature-independent spontaneous emission rate, and time is
in units of $t_0=\hbar/meV$.

Fig. 4. Time-dependent current arising from the electron-hole
recombination $J_{sp}(t)$ for different tunnelling rates of holes
at zero temperature and the applied voltage with duration time
$\Delta s=3~t_0$ and amplitude $\Delta=20$~meV. Current is given
in units of $J_0=2e ~R_{eh}$, where $R_{eh}$ denotes the time and
temperature-independent spontaneous emission rate, and time is in
units of $t_0=\hbar/meV$.

Fig. 5. Time-dependent current arising from the electron-hole
recombination $J_{sp}(t)$ for different temperatures at the
applied voltage with duration time $\Delta s=3~t_0$ and amplitude
$\Delta=20$~meV. Current is given in units of $J_0=2e ~R_{eh}$,
where $R_{eh}$ denotes the time and temperature-independent
spontaneous emission rate, and time is in units of
$t_0=\hbar/meV$.

%\end{small}


\begin{thebibliography}{50}

\bibitem[1]{Ben} O. Benson, C. Santori, M. Pelton, and Y.
Yamamoto, Phys. Rev. Lett. {\bf 84}, (2000) 2513.

\bibitem[2]{Mic} P. Michler et al, Science {\bf 290}, (2000) 2282.

\bibitem[3]{Yua} Z. Yuan et al, Science {\bf 295}, (2002) 102.


\bibitem[4]{Ima} A. Imamoglu, and Y. Yamamoto, Phys. Rev. Lett.
{\bf 72}, (1994) 210.

\bibitem[5]{Han} C. K. Hahn, J. Motohisa and T. Fukui, Appl. Phys.
Lett. \textbf{76}, 3947 (2000) .

\bibitem[6]{Cao} H. Cao, G. Klimovitch, G. Bjork, and Y. Yamamoto,
phys. Rev. B \textbf{52}, 12184 (1995).

\bibitem[7]{Che} Y. N. Chen and D. S. Chuu, Phys. Rev. B \textbf{66},
165316 (2002).

\bibitem[8]{Kuo} D. M. T. Kuo and Y. C. Chang, Phys. Rev. B \textbf{72},
085334 (2005).


\bibitem[9]{Hau} H. Haug and A. P. Jauho, Quantum Kinetics in Transport
and Optics of Semiconductors (Springer, Heidelberg, 1996).

\bibitem[10]{Jau} A. P. Jauho, N. S. Wingreen and Y. Meir, Phys.
Rev. B \textbf{50}, 5528 (1994).

\bibitem[11]{Hau} H. Haug and S. W. koch, "Quantum theory of the
optical and electronic properties of semiconductors", (World
scientific, Singgapor,  1990).

\bibitem[12]{Agu} R. Aguado, J. Inarrea and G. Platero, Phys. Rev.
B \textbf{53}, 10030 (1996).

\bibitem[13]{Sun} Q. F. Sun, J. Wang and T. H. Lin, Phys. Rev. B
\textbf{61}, 12643 (2000).

\bibitem[14]{Mer} J. Merino and J. B. Marston, Phys. Rev. B \textbf{69},
115304 (2004).

\bibitem[15]{Ste} G. Stefanucci and C. O. Almbladh, Phys. Rev. B
\textbf{69}, 195318 (2004).

\bibitem[16]{Wan} B. Wang, J. Wang and H. Guo, Phys. Rev. Lett.
\textbf{82}, 398 (1999).

\bibitem[17]{You} J. Q. You, C. H. Lam and H. Z. Zheng, Phys. Rev.
B \textbf{62}, 1978 (2000).

\bibitem[18]{Win} N. S. Wingreen, A. P. Jauho and Y. Meir, Phys.
Rev. B \textbf{48}, 8487 (1993).

\bibitem[19]{Kuo} David. M. T. Kuo and Y. C. Chang, Phys. Rev. B
\textbf{60}, 15957 (1999).


\bibitem[20]{Sol} G. S. Solomon, M. Pelton and Y. Yamamoto, Phys.
rev. Lett.\textbf{ 86}, 3903 (2001).


\bibitem[21]{Cha} W. H. Chang, W. Y. Chen, H. S. Chang, T. P.
Hsieh, J. I. Chyi and T. M. Hsu, Phys. Rev. Lett. \textbf{96},
117401 (2006).

\bibitem[22] {Eng} D. Englund, D. Fattal, E. Waks, G. Solomon, B.
Zhang, T. Nakaoka, Y. Arakawa, Y. Yamamoto and J. Vuckovic, Phys.
Rev. Lett. \textbf{95}, 013904 (2005).

\bibitem[23] {Ura} J. Urayama, T. B. Norris, J. Singh and P.
Bhattacharya, Phys. Rev. Lett. 86, 4930 (2001).

%\bibitem[25] {Asr} L. V. Asryan and R. A. Suris, Semicond. Sci.
%Technol 11, 554 (1996).





\end{thebibliography}
\end{document}